\documentclass[aps,pra,noshowpacs,twocolumn,preprintnumbers,superscriptaddress,floatfix]{revtex4}

\usepackage{amsmath,amssymb}
\usepackage{multirow}
\usepackage{hyperref,graphicx}
\usepackage{epsfig}
\usepackage{times}
\usepackage{pdfpages}

\begin{document}
\title{Mapping Twisted Light into and out of a Photonic Chip}

\author{Yuan Chen}
\affiliation{State Key Laboratory of Advanced Optical Communication Systems and Networks, School of Physics and Astronomy, Shanghai Jiao Tong University, Shanghai 200240, China}
\affiliation{Institute for Quantum Science and Engineering and Department of Physics, Southern University of Science and Technology, Shenzhen 518055, China}
\affiliation{Synergetic Innovation Center of Quantum Information and Quantum Physics, University of Science and Technology of China, Hefei, Anhui 230026, China}

\author{Jun Gao}
\affiliation{State Key Laboratory of Advanced Optical Communication Systems and Networks, School of Physics and Astronomy, Shanghai Jiao Tong University, Shanghai 200240, China}
\affiliation{Institute for Quantum Science and Engineering and Department of Physics, Southern University of Science and Technology, Shenzhen 518055, China}
\affiliation{Synergetic Innovation Center of Quantum Information and Quantum Physics, University of Science and Technology of China, Hefei, Anhui 230026, China}

\author{Zhi-Qiang Jiao}
\affiliation{State Key Laboratory of Advanced Optical Communication Systems and Networks, School of Physics and Astronomy, Shanghai Jiao Tong University, Shanghai 200240, China}
\affiliation{Synergetic Innovation Center of Quantum Information and Quantum Physics, University of Science and Technology of China, Hefei, Anhui 230026, China}

\author{Ke Sun}
\affiliation{State Key Laboratory of Advanced Optical Communication Systems and Networks, School of Physics and Astronomy, Shanghai Jiao Tong University, Shanghai 200240, China}

\author{Wei-Guan Shen}
\affiliation{State Key Laboratory of Advanced Optical Communication Systems and Networks, School of Physics and Astronomy, Shanghai Jiao Tong University, Shanghai 200240, China}
\affiliation{Synergetic Innovation Center of Quantum Information and Quantum Physics, University of Science and Technology of China, Hefei, Anhui 230026, China}

\author{Lu-Feng Qiao}
\affiliation{State Key Laboratory of Advanced Optical Communication Systems and Networks, School of Physics and Astronomy, Shanghai Jiao Tong University, Shanghai 200240, China}
\affiliation{Synergetic Innovation Center of Quantum Information and Quantum Physics, University of Science and Technology of China, Hefei, Anhui 230026, China}

\author{Hao Tang}
\affiliation{State Key Laboratory of Advanced Optical Communication Systems and Networks, School of Physics and Astronomy, Shanghai Jiao Tong University, Shanghai 200240, China}
\affiliation{Synergetic Innovation Center of Quantum Information and Quantum Physics, University of Science and Technology of China, Hefei, Anhui 230026, China}
\affiliation{Institute of Natural Sciences, Shanghai Jiao Tong University, Shanghai 200240, China}

\author{Xiao-Feng Lin}
\affiliation{State Key Laboratory of Advanced Optical Communication Systems and Networks, School of Physics and Astronomy, Shanghai Jiao Tong University, Shanghai 200240, China}
\affiliation{Synergetic Innovation Center of Quantum Information and Quantum Physics, University of Science and Technology of China, Hefei, Anhui 230026, China}

\author{Xian-Min Jin}
\thanks{xianmin.jin@sjtu.edu.cn}
\affiliation{State Key Laboratory of Advanced Optical Communication Systems and Networks, School of Physics and Astronomy, Shanghai Jiao Tong University, Shanghai 200240, China}
\affiliation{Synergetic Innovation Center of Quantum Information and Quantum Physics, University of Science and Technology of China, Hefei, Anhui 230026, China}
\affiliation{Institute of Natural Sciences, Shanghai Jiao Tong University, Shanghai 200240, China}
\date{\today}

\begin{abstract}
Twisted light carrying orbital angular momentum (OAM) provides an additional degree of freedom for modern optics and an emerging resource for both classical and quantum information technologies. Its inherently infinite dimensions can potentially be exploited by using mode multiplexing to enhance data capacity for sustaining the unprecedented growth in big data and internet traffic, and can be encoded to build large-scale quantum computing machines in high-dimensional Hilbert space. While the emission of twisted light from the surface of integrated devices to free space has been widely investigated, the transmission and processing inside a photonic chip remain to be addressed. Here, we present the first laser-direct-written waveguide being capable of supporting OAM modes and experimentally demonstrate a faithful mapping of twisted light into and out of a photonic chip. The states OAM$_{0}$, OAM$_{-1}$, OAM$_{+1}$ and their superpositions can transmit through the photonic chip with a total efficiency up to $60\%$ with minimal crosstalk. In addition, we present the transmission of quantum twisted light states of single photons and measure the output states with single-photon imaging. Our results may add OAM as a new degree of freedom to be transmitted and manipulated in a photonic chip for high-capacity communication and high-dimensional quantum information processing.
\end{abstract}

\maketitle

The phase of an optical beam with a spatial degree of freedom of OAM is twisted like a corkscrew around its axis of travel and the cancellation of light waves at the axis itself results in a ``doughnut" intensity profile. The twisted light has a helical wave front with an azimuthal phase term $e^{ i \ell \varphi}$ \cite{Allen1992}, with which every photon can carry an OAM of  ${\ell\hbar}$ ( ${\ell}$ is topological charge, ${\varphi}$ is azimuthal angle, and ${\hbar}$ is Planck constant ${\it h}$ divided by ${2\pi}$). Having the special features of intensity structure (``doughnut"  intensity), phase structure (spiral phase front) and dynamic characteristic (carrying OAM), the twisted light has been widely applied into the field of optical manipulation \cite{Dholakia2011}, optical trapping \cite{Paterson2001,MacDonald2002} and optical tweezers \cite{Padgett2011}.

In recent years, OAM has shown the great potential in communication systems to overcome the channel capacity crunch \cite{Wang2012,Bozinovic2013}. The unlimited topological charges and the inherent orthogonality may provide tremendous resources for mode multiplexing. The inherently infinite dimensions potentially can also be exploited to deliver high-dimensional quantum states with larger alphabets and to build quantum computing machines in high-dimensional Hilbert space \cite{Dada2011,Fickler2012,Krenn2014,Mirhosseini2015,Bouchard2017}.

Large-scale applications of OAM beyond proof-of-principle demonstrations require developing integrated devices to enable the generation, transmission and processing of such a new degree of freedom. Previous works have demonstrated on-chip generation twisted light with integrated star couplers \cite{Doerr2011}, micro-ring resonators \cite{Cai2012} and controlled phase arrays \cite{Sun2014}. While the emission of twisted light from the surface of integrated devices to free space has been widely investigated, the transmission and processing inside a photonic chip remain to be solved. 

In this letter, we demonstrate a faithful mapping of twisted light into and out of a photonic chip by prototyping ``doughnut" waveguides with femtosecond laser direct writing \cite{Rafael2008,Szameit2010}. We couple the states OAM$_{0}$, OAM$_{-1}$, OAM$_{+1}$ and their superpositions into and out of the photonic chip with a total efficiency up to $60\%$ and verify the output states by interfering with Gaussian reference beam and making projection measurements, which clearly demonstrate that the output state basically keeps the OAM of the input state. In addition, we present the transmission of single-photon quantum twisted light and measure the output states with single-photon imaging.

\begin{figure*}[htbp!]
\centering
\includegraphics[width=1.56\columnwidth]{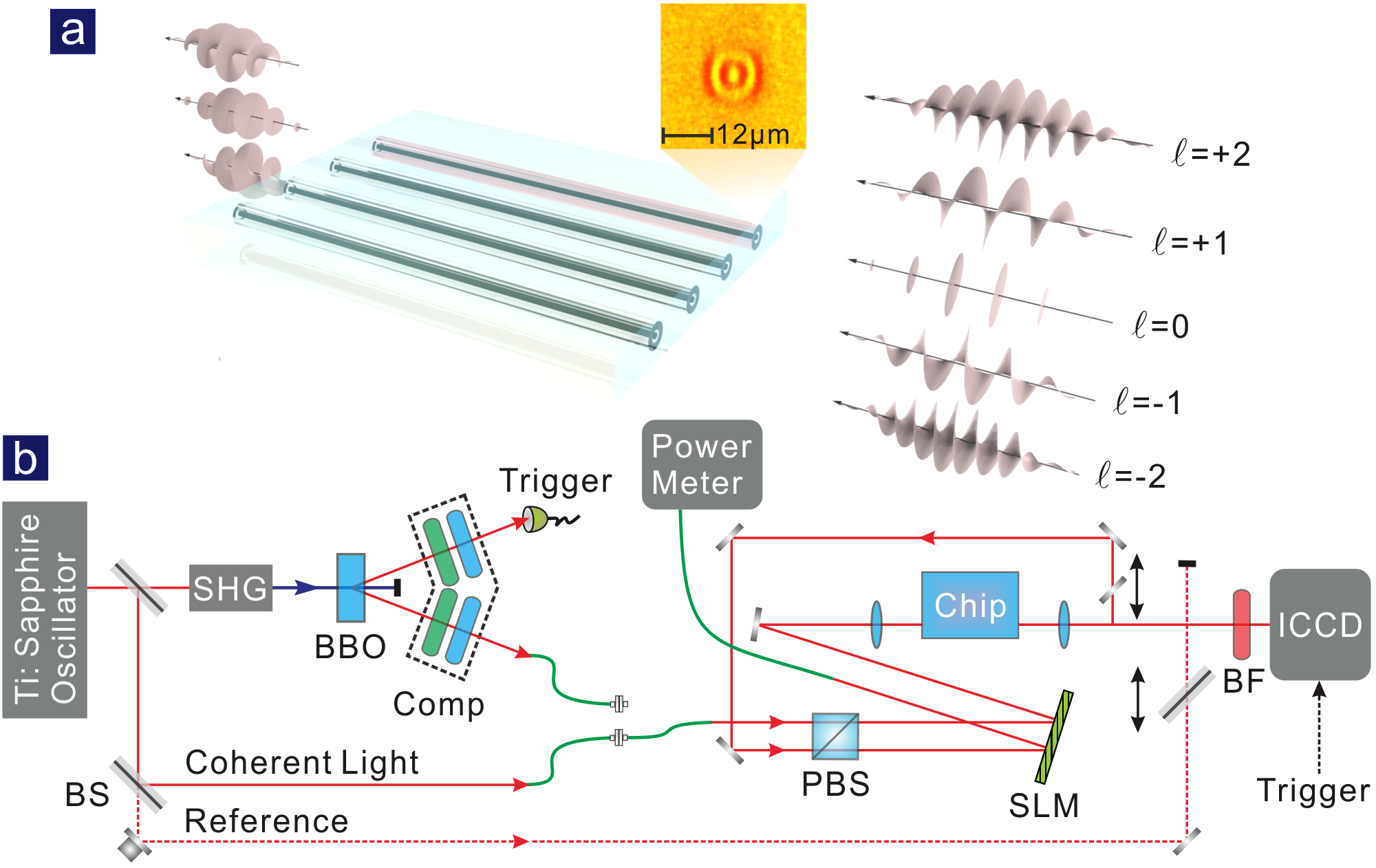}
\caption{\textbf{Experimental implementation.} \textbf{a.} Schematic diagram of mapping twisted light into and out of a photonic chip. The inset shows the cross-section image of a femtosecond laser-written ``doughnut" waveguide. The OAM modes are generated externally, then coupled into ``doughnut" waveguide, and finally analyzed after chip. \textbf{b.} The experimental setup with classical and quantum probes in OAM degree of freedom. Half of the SLM is used to generate the twisted light, and the other half is used to make projection measurement for the output states via a standard phase-flattening method. By switching the holograms, we can observe the image in the far field. When the hologram applied to the SLM is opposite to the OAM mode, the far field will be a Gaussian mode, representing an effective projection. We select the Gaussian component, and couple the light into a single-mode fiber (green line) to measure the OAM spectra after the transmission by using a power meter. Comp: half wave plate and quarter wave plate together are used to polarization compensation; BS: beam splitter; PBS: polarization beam splitter; SLM: spatial light modulator; BF: band-pass filter (780 nm $\pm$ 6 nm); ICCD: intensified charge coupled device camera.}
\label{Figure 1}
\end{figure*}

In conventional waveguides, the effective index $\it n_{eff}$ is much too small to isolate near-degenerate OAM states from one another. A solution to this problem is to enhance the vector splitting through a choice of waveguide structure. It is known that the typical transverse intensity pattern of a OAM beam is ``doughnut" shaped. The waveguides with such shaped cross-section, being cylindrically symmetric structure, may support OAM modes. The key physical problem of OAM propagating in cylindrically symmetric structure is that, during total internal reflection, phase shifts at index discontinuities critically depend on polarization orientation of an incident wave \cite{Ramachandran2013}. To translate this physical problem into a mathematical problem, it can be expressed by a full vectorial solution of the Maxwell equations. By means of a first-order perturbative analysis, it is found that a ``doughnut" waveguide would be more suitable for supporting OAM modes \cite{Ramachandran2013}. 

We realize constructing such a three-dimensional (3D) structure by using femtosecond laser direct writing technique (see Supplemental Material A \cite{SectionA,Osellame2012}). Fig. 1a inset shows the cross-section image of the fabricated waveguide, which is measured by an optical microscope with a built-in Kohler illumination system (a halogen lamp light source). We prepare the twisted light in different OAM modes and couple them into the ``doughnut" waveguide embeded in the photonic chip and expect to observe well-preserved output states. After polishing both ends of the chip, the length of ``doughnut" waveguide is 19.64mm. As is shown in Fig. 1b, the probe is switchable to both coherent light and heralded single photon (see Supplemental Material B \cite{SectionB}). In our experiment, the coupling objective is optimized for mode matching of OAM$_{-1}$, OAM$_{0}$, OAM$_{+1}$ between free-space beam profile and waveguide cross-section. After mapping out of the photonic chip, we measure the intensity profile with a CCD camera. The total efficiency is obtained by measuring the probe power before and after the chip, which is up to $60.0\%$. 

\begin{figure*}[htb!]
\centering
\includegraphics[width=2.02\columnwidth]{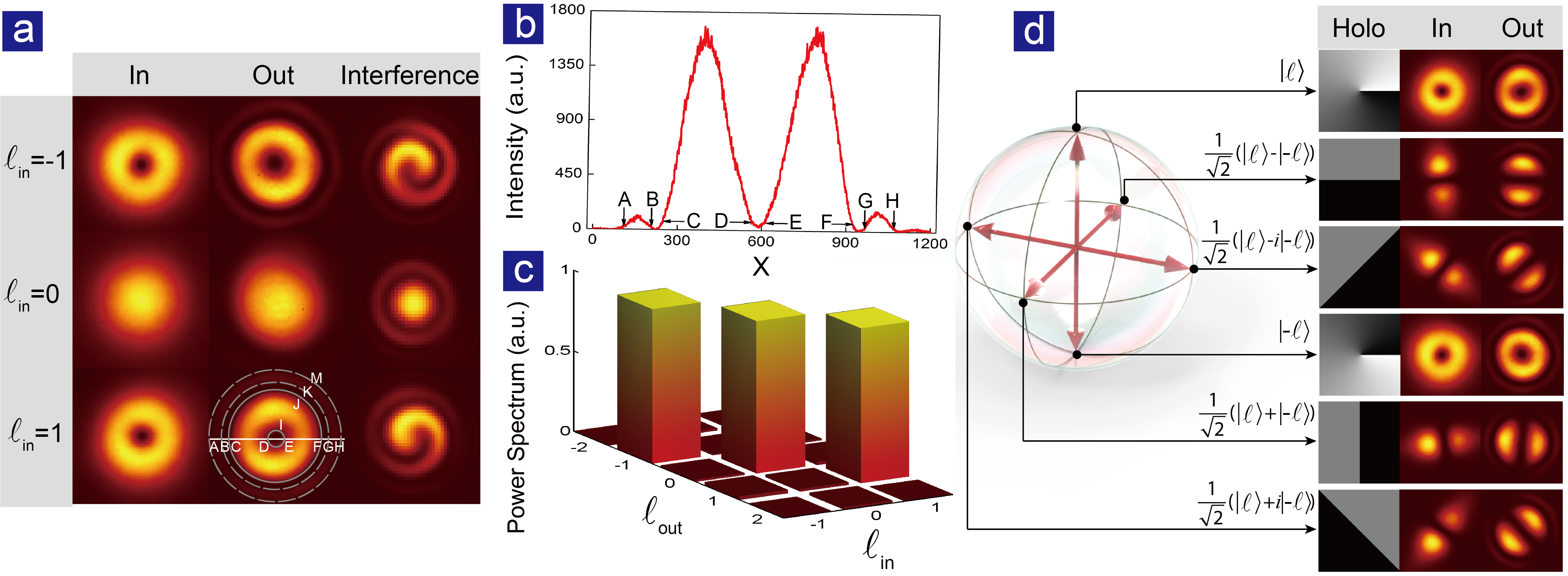}
\caption{\textbf{Experimental results of mapping OAM$_{-1}$, OAM$_{0}$, OAM$_{+1}$ and their superpositions}. \textbf{a.} The measured intensity profiles of OAM$_{-1}$, OAM$_{0}$, OAM$_{+1}$ modes before (after) the chip are shown in the first (second) column. Measured interference patterns shown in third column clearly confirmed faithful preservation of OAM. \textbf{b.} Radial intensity distribution extracted from the OAM$_{+1}$ mode along the white radial direction. \textbf{c.} Measured OAM power spectra after the chip. \textbf{d.} Experimental results of OAM superposition states for ${\ell}=1$. The holograms applied to SLM and the measured intensity profiles for six universal states are presented visually pointing to Bloch sphere.}
\label{Figure 2}
\end{figure*}

The observed intensity profiles of OAM$_{-1}$, OAM$_{0}$, OAM$_{+1}$ modes are shown in Fig. 2a. The prepared states before mapping into the chip are shown in the first column and the output states are shown in the second column. By inserting a beamsplitter, we are able to measure the interference with a Gaussian reference beam. The yielded interference pattern shown in the third column of Fig. 2a can be employed to verify the topological charge of the output OAM modes. The high-visibility clockwise (counterclockwise) spiral interference patterns are observed for OAM$_{+1}$ (OAM$_{-1}$) output state, which indicates a great ability to maintain the twisted light.

In order to analyze the output states quantitatively, we take OAM$_{+1}$ mode after chip as a typical example to analyze the intensity ratio between the outer ring and inner ring. A, B, G and H (C, D, E and F) marked in Fig. 2a are the junction points of the outer (inner) ring with the white radial direction. By a trapezoidal integration on the 1D radial intensity distribution shown in Fig. 2b, we multiply it by the ratio of lengths of the two circles (about a factor of $2.5$), and then we can obtain the power of outer ring, which is about $7.5\%$ of the inner ring power. More rigorously, we make an integration on the cross-area of the rings. We mark the inner (outer) ring with the two solid (dotted) circles I and J (K and M). The obtained power of the outer ring is about $6.7\%$ that of the inner ring. This indicates that the radial index has a limited influence on the output state. We therefore focus on the azimuth information of OAM modes. As is shown in Fig. 2c, we make projection measurements (see Supplemental Material C \cite{SectionC}) for the output states and the corresponding percentage of power that remains on the same OAM value after going through the waveguide is over $95.5\%$ on average, which indicates again that the twisted light is well maintained through the chip.

Besides individual pure states, we further look into the ability to support general superposition states. We can construct matrices to describe the superpositions of OAM$_{\pm|{\ell}|}$ and their transformations on two-dimensional subspaces \cite{Allen1999} that can be represented by a Bloch sphere, equivalent to the Poincar\'e sphere for polarization \cite{Padgett1999}. By tuning the relative phase and the magnitude of the superposition components on a spatial light modulator, we can access states distributed over the entire Bloch sphere \cite{Jack2010}. We show the results of mapping six universal states on Bloch sphere into and out of the photonic chip, which indicate great capacity to simultaneously support all superposition states (see Fig. 2d). Interestingly, we also observe that output states possess even better spatial structure than the input states. It can be understood that the unwanted components of the imperfect input states can be spatially filtered out by the waveguide that only supports well-defined OAM modes. We attribute the unwanted components generated with the SLM to a convergent Gaussian spot imprinted on the SLM. In addition, an equal-weighted superposition state consisting of OAM$_{0}$, OAM$_{-1}$, OAM$_{+1}$ can also be transmitted through the chip (see Supplemental Material D \cite{SectionD}), which is a promising hint, that our doughnut-shaped waveguide can support genuinely high-dimensional states in very large spaces.

Our photonic chip is optimized for supporting OAM$_{0}$, OAM$_{-1}$, OAM$_{+1}$ and their superpositions. It would be interesting to explore the propagation properties of higher-order modes. As is shown in Fig. 3a, while the sign of positive and negative topological charges are unchanged, the states are all mapped onto their corresponding first-order OAM modes, which are clearly revealed by the chirality and the number of arms in the measured interference patterns.

To understand this, we have also made projection measurement on the input and output states for all the higher-order OAM modes. As is shown in Fig. 3b, the measured power spectra verifies that the input states for all the higher-order modes are almost pure states with an average purity up to $94.2\%$, with a negligible component of $0.4\%$ ($0.5\%$) for OAM$_{+1}$ (OAM$_{-1}$). However, the measured power spectra of the output states are found to mainly weight on OAM$_{-1}$ or OAM$_{+1}$ mode depending on the chirality of the input states (see Fig. 3c). In particular, for the negative (positive) higher-order input modes, the measured weight of the output states on OAM$_{-1}$ (OAM$_{+1}$) is dominant, reaching $66.9\%$ ($61.2\%$) on average. These results explain why we can only observe the interference pattern with one spiral arm and chirality. 

\begin{figure*}[htb!]
\centering
\includegraphics[width=1.6\columnwidth]{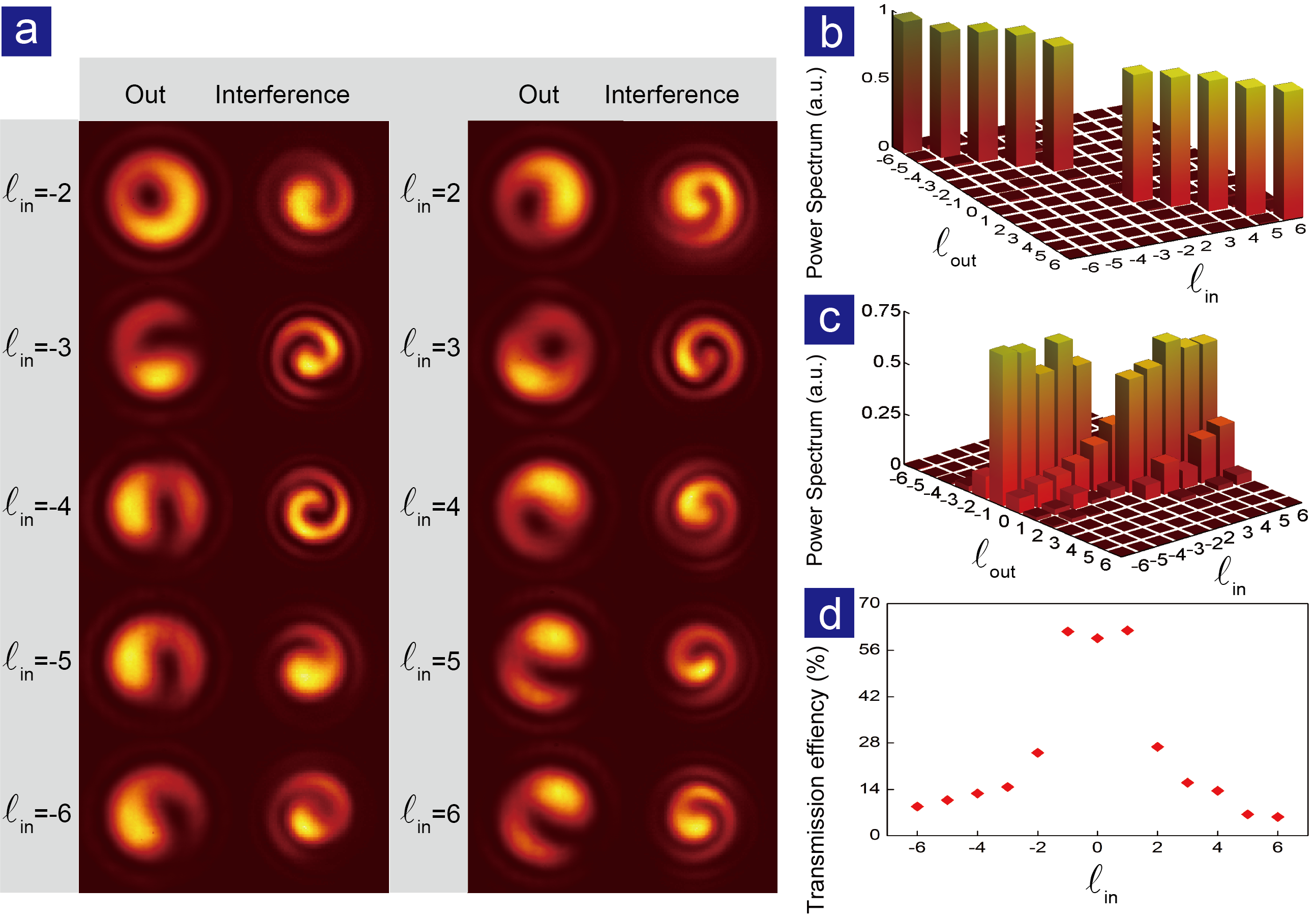}
\caption{\textbf{Experimental mapping of higher-order OAM modes}. \textbf{a.} The measured intensity profiles and interference pattern on the output end of the chip for high-order ${\ell}_{\rm in}$ up to 6. \textbf{b.} Measured OAM power spectra before the chip. \textbf{c.} Corresponding OAM power spectra after the chip. \textbf{d.} Measured transmission effiency versus different input states.}
\label{Figure 3}
\end{figure*}

\begin{figure}[htb!]
\centering
\includegraphics[width=1.0\columnwidth]{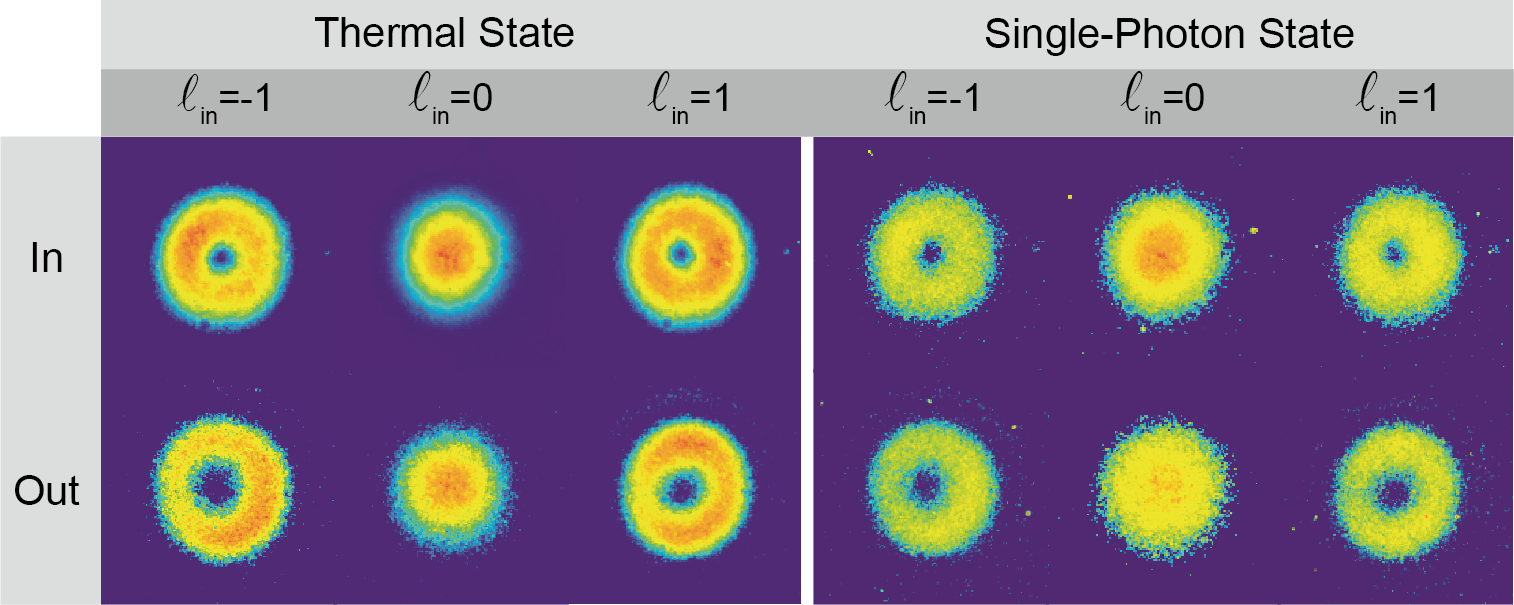}
\caption{\textbf{Experimental results of OAM quantum states}. The intensity profiles of quantum twisted light in heralded single-photon states (thermal states) are measured directly with (without) externally triggering the ICCD with another photon. The single-photon images are shown without any reduction of background noises.}
\label{Figure 4}
\end{figure}

We further measure the transmission efficiency for all the input OAM modes and observe a dramatic drop from the first-order ($61.8\%$) to the second-order ($25.9\%$) input OAM modes (see Fig. 3d), while the transmission efficiency drops gradually from the second-order to the sixth-order input OAM modes. Therefore, when higher-order OAM mode is coupled, it will dissipate and partially evolve into the eigen mode, as the output is always mainly OAM$_{-1}$ (OAM$_{+1}$). These results imply the ``doughnut" waveguide has its own eigen modes OAM$_{0}$, OAM$_{-1}$ and OAM$_{+1}$ (see Supplemental Material E \cite{SectionE,Curtis2003}) and tends to support the twisted light of lower orders, which further contributes to the observed conversion effect \cite{Yang2017}. 

The inherently infinite dimensions potentially can be exploited as an alternative resource to prepare high-dimensional Hilbert space quantum states, hyper-entanglement for instance, to boost the computational power of quantum computing and quantum simulations, rather than to prepare quantum states with higher photon number. The resource of photonic dimensions seems more scalable than photon number \cite{Fickler2016}, but requires complex processing circuits and phase-level stability, i.e. is physically unscalable with bulk optics. Quantum integrated photonics would be an elegant solution for such demands. We make the step forward for OAM-based high-dimensional quantum information processing by demonstrating mapping single-photon quantum twisted light into and out of a photonic chip. Thanks to the rapid progress in imaging technologies over the last few years, CCD cameras has become an interesting option for single-photon detection in quantum optics experiments, since the large spatial information \cite{Fickler2013, Ding2013} is directly accessible. 

In this experiment, we directly visualize quantum twisted light in thermal states and heralded single-photon states before and after the photonic chip by employing an ICCD camera. We initialize (or spatially filter) the thermal states and heralded single-photon states first with a single-mode fiber before imprinting OAM, which means the photons are all coherent in transverse spatial domain. The obtained intensity profiles are shown in Fig. 4 without any Fourier-transformation-based noise filtering and background reduction. It should be noticed that the high-quality imaging of heralded single-photon states are achieved because we utilize the detection signal of another photon to trigger ICCD with a time window as narrow as 10 $ns$. 

In summary, we demonstrate a faithful and highly efficient mapping of twisted light into and out of a photonic chip by prototyping ``doughnut" waveguides with femtosecond laser direct writing. The simultaneous support of the states OAM$_{-1}$, OAM$_{0}$, OAM$_{+1}$ and their superpositions suggests that it is possible to transmit and manipulate OAM states inside a photonic chip. We also show the compatiblity of single-photon quantum twisted light to the photonic chip and measure the output states with single-photon imaging, which may promise OAM based integrated high-dimensional quantum information processing. 

This emerging field of twisted-light-inside integrated photonics has many open problems to be solved. The evanescent light coupling or splitting between two OAM waveguides is a primary goal, which may facilitate the design and fabrication of many novel OAM carrying integrated devices, especially may enable OAM state generation inside chip by appropriate coupling and phase matching \cite{Mohanty2017}. Quantum interference between the transverse spatial modes has been observed in multimode waveguide \cite{Mohanty2017}, which implies that it possible to realize the first-order or Hong-Ou-Mandel interference for OAM modes inside a photonic chip. 

Multi-channel all-on-chip sender (receiver) can be conceived to encode (sort) OAM with large alphabet information \cite{Bechmann-Pasquinucci2000}, for both classical and quantum communications. The combined use of different degrees of freedom of a single photon, such as spin and orbital angular momentum, enables the on-chip implementation of entirely new quantum information systems in a high-dimensional space \cite{Bouchard2017,Krenn2017,Erhard2018} for quantum supremacy.

\section*{Acknowledgements} The authors thank Miles Padgett, Nathan Langford and Jian-Wei Pan for helpful discussions and suggestions. This work was supported by National Key R\&D Program of China (2017YFA0303700); National Natural Science Foundation of China (NSFC) (61734005, 11761141014, 11690033); Science and Technology Commission of Shanghai Municipality (STCSM) (15QA1402200, 16JC1400405, 17JC1400403); Shanghai Municipal Education Commission (SMEC)(16SG09, 2017-01-07-00-02-E00049); X.-M.J. acknowledges support from the National Young 1000 Talents Plan.

\section*{Supplementary Material:}
\subsection{\textbf{3D Fabrication of ``doughnut" waveguides}}
The ``doughnut" waveguide proposed to support OAM modes in a photonic chip is very different from using a single-mode fiber. Its cylindrically symmetric structure requires 3D fabrication capacity, which is very challenging to be realized with conventional fabrication methods of silicon photonics. We employ femtosecond laser direct writing technique to realize such 3D fabrication capacity. Due to nonlinear absorption effects, the wafer materials only absorb energy in a scope of micrometer level and a short time slot of hundreds of femtoseconds, and therefore the refractive index inside the wafer can be modified in a very small scale \cite{Rafael2008,Szameit2010,Osellame2012} around the laser focal spot. Continuous scanning the wafer and/or the laser focal spot will allow us to manufacture a very thin line in three dimensions. Multiple writing in such way can construct the proposed ``doughnut" structure piece by piece, just like a ``surgery operation". 

\renewcommand{\thefigure}{\arabic{figure}}
\begin{figure}[htb!]
 \centering
 \includegraphics[width=1\columnwidth]{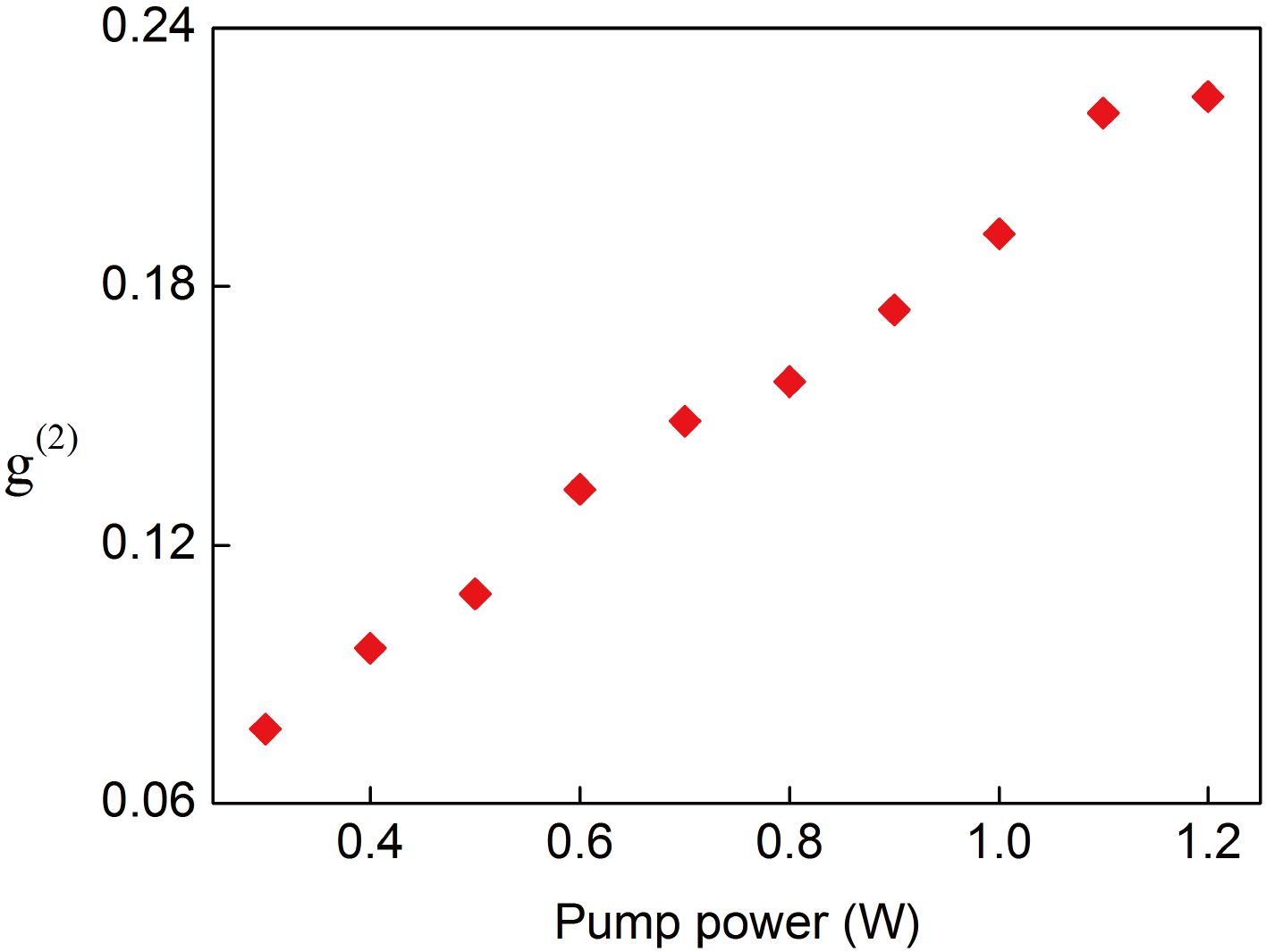}\\
 \caption{
  \textbf{The measured ${\rm g}^{(2)}$ versus pump power.}}
 \label{Figure S1}
\end{figure}

Twelve written waveguides constitute the annular structure, of which the diameter of the circle formed by the core positions of the twelve waveguides is set at 8 $\mu$m. These twelve written waveguides are overlapped to form a continuous refractive index distribution. The central angle of the two adjacent waveguides is 30 degrees. So the linear distance between the two adjacent waveguides is 2.09 $\mu$m. The diameter of single waveguide is about 2.5 $\mu$m. Therefore, the overlap between two adjacent waveguides is estimated at 0.4 $\mu$m. The ``doughnut" structure will consist of 13 waveguides when we apply an additional scan through the middle. The femtosecond laser pulses, with a wavelength of 513 nm, a pulse duration of 290 $fs$ and a repetition rate of 1 MHz, are focused into the volume of borosilicate glass wafer (EAGLE 2000, Corning Inc.) by a 0.7 numerical aperture microscope objective. The wafer size is a 1$\times$20$\times$20 mm. Under suitable irradiation conditions (60 $nJ$ pulse energy and 5 mm/s translation speed), waveguides are produced at an average depth of 170 $\mu$m underneath the glass surface using 3-axis air-bearing stages (Aerotech Inc.). The refractive index contrast and birefringence are estimated at the order of $10^{-3}$ and $10^{-5}$, respectively.

\subsection{\textbf{Classical and quantum twisted light preparation}} As is shown in Fig. 1b, a Ti: Sapphire Oscillator centred at 780 nm is divided into three beams by inserting two beamsplitters. One of the beams is relatively weak and serves as reference to measure the interference patterns. A translation stage is added in order to tune the phase for high-contrast interference fringes. The second beam as coherent light, weak as well, is utilized to prepare classical twisted light. The third beam, the strongest one, is employed to produce a 390 nm laser up to 1.2 W via second harmonic generation (SHG). We feed the up-converted laser into a 2-mm-thick BBO crystal tuned for type II, non-collinear down-conversion to prepare a photon pair. The obtained single-channel count rate and two-channel coincidence count rate are 1875000 and 237500 respectively. 

\renewcommand{\thefigure}{\arabic{figure}}
\begin{figure}[htb!]
\centering
\includegraphics[width=1\columnwidth]{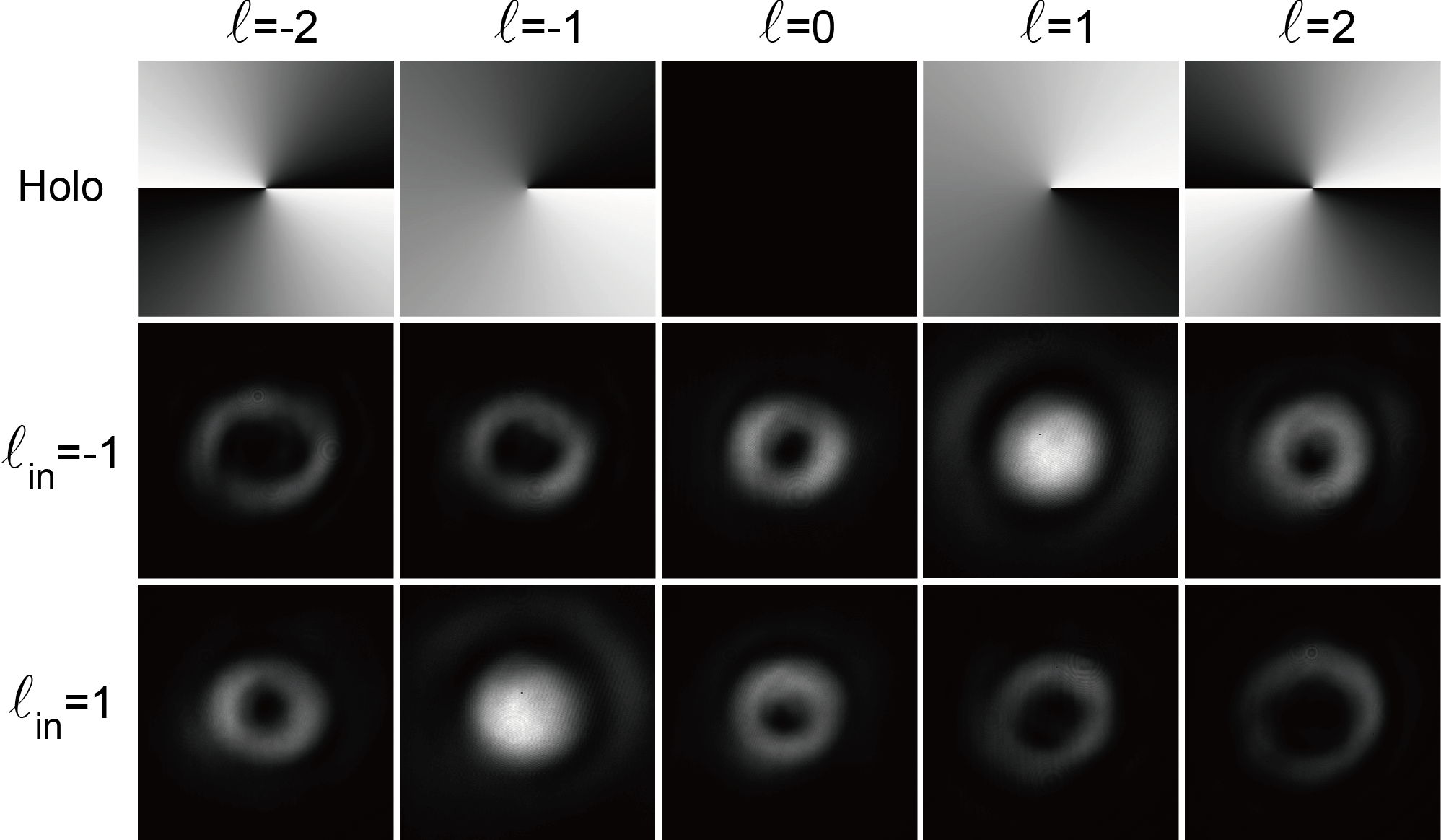}
\caption{\textbf{The image in the far field for power spectrum measurement}. The first row is the hologram applied to SLM for projection measurement; the second (third) row is the intensity profile obtained after projection measurement corresponding to the first row hologram for OAM$_{-1}$ or OAM$_{+1}$ mode.}
\label{Figure S2}
\end{figure}

In our experiment, we initialize (or spatially filter) the thermal states and heralded single photons states first with a single-mode fiber before imprinting OAM. It means the photons are perfect coherent in the transverse spatial domain.
The concepts of thermal in photon statistics and coherent in transverse spatial domain are very different and sometime are quite confusing. The two-mode squeezed states generated by SPDC source can be described by $\vert{\psi}\rangle=\frac{1}{\sqrt{1-\lambda}}[\vert{0_{A}}{0_{B}}\rangle +\lambda\vert{1_{A}}{1_{B}}\rangle+\lambda^2\vert{2_{A}}{2_{B}}\rangle+\cdots]$, where $\lambda$ is nonlinear coefficient. When we trace the arm B out, the reduced density operator can be expressed as $\rho_{A}=\frac{1}{1-\lambda}[\vert{0}\rangle_{A}\langle{0}\vert +\lambda^2\vert{1}\rangle_{A}\langle{1}\vert+\lambda^4\vert{2}\rangle_{A}\langle{2}\vert+\cdots]$. The statistical distribution of photon number of the nonheralded single photons shows a super-Poissonian distribution.

One photon registered at an avalanched photo diode can herald the existence of a well-defined single photon. We measured the single photoness of our source and the result is shown in Fig. 5. There is a tradeoff between $\lambda$ and single photoness. In our experiment, we scan the pump power from 0.3 W to 1.2 W, and finally take our data at 1.2 W, with which we can have a reasonable good single photons and high photon pair rate. The heralding signal is sent to ICCD to open a window of 10 $ns$ for taking the image of the heralded single-photon OAM state. We use a 6m fiber as a delay line to image the single photons. And in the external trigger mode, the insertion delay of ICCD is set as 35ns. Considering the delay induced by the optical path in our setup, the delay of 6-meter-long fiber is enough to open the gate beforehand. The single photon and coherent light can be switched easily by fiber flanges to an individual set-up for converting single-mode beam to the twisted light. The total efficiency of the prepared classical and quantum twisted light can reach 60\%. 

\begin{figure}[htb!]
\centering
\includegraphics[width=0.9\columnwidth]{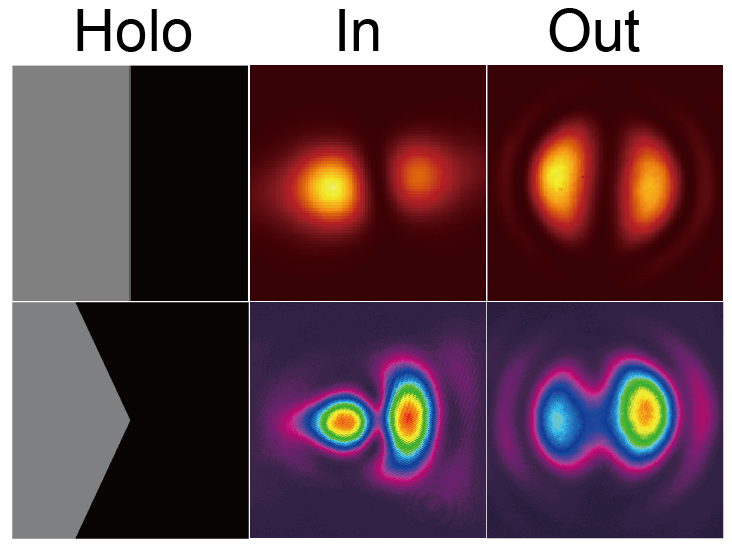}
\caption{\textbf{The result of high-dimensional superposition states}.The hologram and measured intensity profiles for two-dimensional and three-dimensional superposition states before and after the chip.}
\label{Figure S2}
\end{figure}

\subsection{\textbf{Power spectrum measurement}} To analyze the output states quantitatively, we use the SLM for both twisted light generation and projection, half of the SLM is used to generate the input states, and the other half is used to make projection measurement for the output states. Firstly, we employ the output Gaussian mode from the chip to align with the singularity of the hologram used for projection. Second, we change to the OAM modes and switch the holograms, we can observe the images (see Fig. 6) in the far field. When the order of the hologram on SLM is opposite to the output state, resulting in planar phase fronts. Third, we employ SLM to project and select the Gaussian component, and then couple the light into a single-mode fiber to measure the OAM spectra after the transmission. This is a simple phase flattening method, which is a good approximation. For the higher-order input OAM modes, when we switch the holograms, we can also observe the images in the far field. We find that when the output states projected on OAM$_{+1}$ or OAM$_{-1}$, the resulting image in the far field is a Gaussian component. And then we couple the light into a single mode fiber to only select the Gaussian component. 

\begin{figure*}[htb!]
\centering
\includegraphics[width=2\columnwidth]{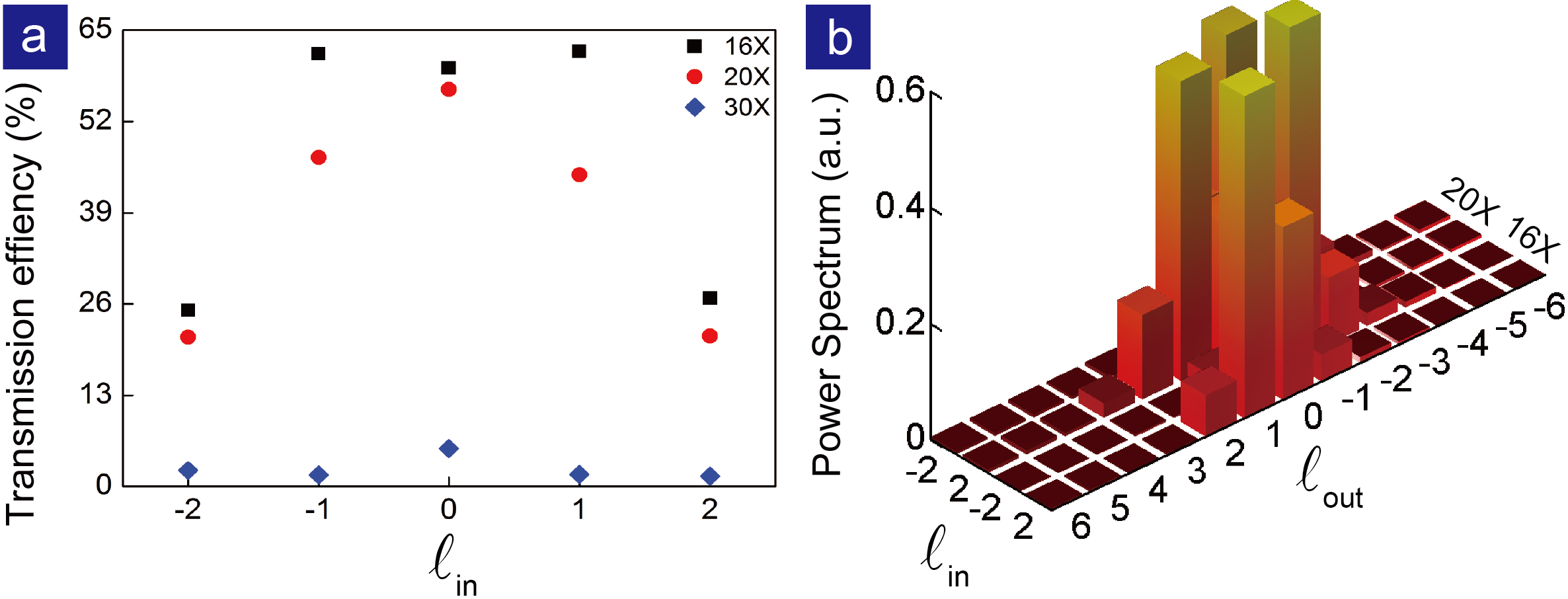}
\caption{\textbf{The effect of different coupling system on the results}. \textbf{a.}Measured transmission effiency versus different input states with different coupling system. \textbf{b.} Measured OAM power spectra after the chip with 20X (16X) coupling system for second-order OAM modes.The output state is mainly weighted on OAM$_{-1}$ or OAM$_{+1}$ mode depending on the chirality of input state no matter what the coupling system is 20X or 16X.}
\label{Figure S2}
\end{figure*}

\subsection{\textbf{High-dimensional superposition state}} We further explore the the performance on high-dimensional superposition state, which is an equal-weighted superposition state consisting of OAM$_{0}$, OAM$_{-1}$, OAM$_{+1}$. As is shown in Fig. 7, compared with two-dimensional superposition state of OAM$_{-1}$ and OAM$_{+1}$, three-dimensional superposition states are obviously uneven two-lobes with stickiness, which indicates that the output states can well preserve the intensity profile. The transmission efficiency of three-dimensional superposition state decreases by 3.5\% compared with two-dimensional superposition state. Therefore, we can see that the three-dimension superposition state can be well preserved, which indicates that our waveguide structure would support high-dimensional states of light. 

\subsection{\textbf{The effect of different coupling system on the results}} The chip or the current coupling system (16X) is optimized for OAM$_{0}$, OAM$_{-1}$, OAM$_{+1}$ modes. According to Ref \cite{Curtis2003}, the diameter of maximum optical intensity in the focal plane for OAM modes would tend to scale linearly with the topological charge. We further consider whether focusing tighter of the coupling system will make a difference, trying to focus the higher-order OAM modes tighter to spatially match the 8$\mu$m diameter. The results are shown in Fig. 8. Compared with 16X coupling system, the transmission efficiency for first (second)-order OAM mode drops about 16.1\% (4.6\%) on average with a 20X coupling system. With a 30X coupling system, the transmission efficiency drops sharply to less than 5\% for all the first (second)-order OAM modes. The parameter of different coupling system is shown in Table I. With the 20X coupling system, we also make projection measurements for second-order input modes. We can see that the output state is still mainly weighted on OAM$_{-1}$ or OAM$_{+1}$ mode. The results imply that the eigen mode of this ``doughnut" waveguide is OAM$_{0}$, OAM$_{-1}$, OAM$_{+1}$ mode, and higher-order input modes does not match the waveguide. Therefore, when higher-order OAM mode is coupled, it will dissipate and partially evolve into the eigen mode, as the output is always mainly OAM$_{-1}$ (OAM$_{+1}$). Even with tighter focus, the current ``doughnut" waveguide with 8$\mu$m diameter do not support higher-order OAM modes.

\begin{table}[htbp!]
\centering  
\caption{The parameter of different coupling system.}
\setlength{\tabcolsep}{5mm}{
\begin{tabular}{|l|c|c|r} 
\hline
\ Lens& Focal Length & N.A. \\ \hline  
16X &11.0 mm & 0.25 \\   \hline     
20X &8.0 mm & 0.50 \\   \hline     
30X &6.2 mm & 0.40  \\   \hline
\end{tabular}}
\end{table}

\end{document}